\documentclass[a4paper,twocolumn,]{emulateapj}
\slugcomment{}

\usepackage{epsfig,wrapfig,color,sidecap}
\usepackage{index}
\usepackage{subfigure}

\usepackage{colordvi}

\usepackage[usenames,dvipsnames]{xcolor}

\usepackage{times}
\usepackage{ulem}
\usepackage[utf8x]{inputenc}
\usepackage{mathrsfs}

\usepackage{multirow}

\newlength{\figwidth}
\shorttitle{}
\shortauthors{Benyamin et al.}

\RequirePackage[colorlinks=true
,urlcolor=blue
,anchorcolor=blue
,citecolor=blue
,filecolor=blue
,linkcolor=blue
,menucolor=blue
,pagecolor=blue
,linktocpage=true
,pdfproducer=medialab
]{hyperref}

\begin{document}

\bibliographystyle{authordate1}

\title{
The  B/C and sub-Iron/Iron Cosmic ray ratios -  further evidence in favor of the spiral arm  diffusion model}

\author{David Benyamin$^1$,Ehud Nakar$^2$, Tsvi Piran$^1$ \& Nir J. Shaviv$^{1,3}$} 
\affil{1. The Racah Institute of physics, The Hebrew University of Jerusalem, Jerusalem 91904, Israel}
\affil{2. Raymond and Beverly Sackler School of Physics \& Astronomy, Tel Aviv University, Tel Aviv 69978, Israel}
\affil{3. School of Natural Sciences, Institute for Advanced Study, Einstein Drive, Princeton, NJ 08540}

\begin{abstract}

The Boron to Carbon (B/C) and sub-Fe/Fe ratios provides an important clue on Cosmic Ray (CR) propagation within the Galaxy. These ratios estimate the grammage that the CR traverse as they propagate from their sources to Earth.  Attempts to explain these ratios within the standard CR propagation models require ad hoc modifications and even with those  these models necessitate inconsistent grammages to explain both ratios. As an alternative, physically motivated model, we have proposed that CR originate preferably within the galactic spiral arms.  CR propagation from dynamic spiral arms has important imprints on various secondary to primary ratios, such as the B/C ratio and the positron fraction. We use our spiral arm diffusion model with the spallation network extended up to Nickel to calculate the sub-Fe/Fe ratio. We show that without any additional parameters  the spiral arm model consistently explains both ratios with the  same grammage, providing further evidence in favor of this model.

\end{abstract}


\section{introduction}
\label{sec:intro}
\maketitle

Cosmic ray (CR) composition measurements have been collected for almost four decades. Their analysis provides information on the propagation through the interstellar medium (ISM), which further provides information on the CR sources and various properties of the ISM. One of the most important aspects characterizing the propagation is the CR path length distribution (PLD), which describes the probability distribution function of the path lengths traversed by CRs between their origin and their measurement near Earth. This PLD determines the amount of spallation and radioactive decay the particles will undergo on the way to Earth. Thus, by measuring the ratio between secondary CR particles, formed en route, to primary particles accelerated at the source, one can constrain the PLD. 

Until recently one of the stringent and often implicit assumptions in CR propagation models was an azimuthally symmetric source distribution \citep[e.g.][]{StrongReview,Dragon}. The PLD resulting from such models closely resembles an exponential PLD. 

One of the problems in these disk-like models  is the amount of grammage the models require to explain the B/C and the sub-Fe/Fe ratios given the same geometric parameters. Because of the higher spallation cross-sections, the sub-Fe/Fe ratio is typically sensitive to the relative fraction of short path lengths in the PLD. The B/C ratio, with lower spallations cross sections, is more sensitive to the relative fraction of long path lengths. When trying to fit both measured ratios simultaneously in a disk-like model, one finds inconsistency with the nearly exponential PLD of the disk-like models---the sub-Fe/Fe ratio requires a larger mean grammage than the B/C ratio. \cite{GarciaMunoz} suggested that this inconsistency could be resolved with the ad hoc assumption that short path lengths are suppressed.

In a different analysis, \citet{Davis}  investigated whether the introduction of a second, low-energy CR source can reproduce the B/C and sub-Fe/Fe ratios, in particular, at low energies. Although they found that one can recover the B/C ratio and the sub-Fe/Fe ratio separately, they were unable to find any combination of parameters that fit both ratios simultaneously, for the aforementioned reason. Namely, Iron nuclei do not produce enough secondaries given model parameters that correctly reproduce the B/C. 

 Over the past decade, CRs propagation models evolved to include spiral arms as the source of the CR particles \citep{ShavivNewAstronomy,Pamela,DRAGONspiral,SpiralModel,GALPROPspiral}.
In these models,  the CR sources primarily reside at a finite distance from the solar system, and this  naturally suppresses the short path lengths (e.g., see figs.\ 4 and 6 in \citealt{BoverC}, hereafter B14). Hence, spiral arm models could in principle fit both ratios with the same model parameters. We study this possibility here.

B14 considered the spiral arms to be dynamic as well, showing that this model  recovers the low energy behavior of the B/C ratio.
More specifically, secondaries to primaries ratios such as the B/C, are expected to decrease with energy because the time required for particles to reach the solar system (i.e., their ``age") decreases with the diffusivity, which increases with the energy. However, below 1~GeV/nuc., the B/C ratio exhibits the opposite behavior. Although unexpected from just diffusion, it can be explained by taking into account the dynamics of the spiral arms. This is because
 at sufficiently low energies the CR age will be governed not by the diffusion from the spiral arms but instead by the time since the last spiral arm passage, which would be shorter. Since CRs below 1~GeV/nuc. are non-relativistic, an energy independent propagation time then gives less spallation products at lower energies, thus reproducing the B/C rise at low energies (B14), without requiring any additional assumptions on the propagations (such as a galactic wind, reaccaleration or breaks in the diffusion coefficient). 

One interesting ramification of the modified PLD of the spiral arm model is that it requires a notably smaller halo size and diffusion coefficients to recover the observed B/C and $^{10}$Be$/$Be ratio. Namely, by modifying the PLD one has to change the ``canonical" diffusion parameters characterizing the cosmic rays (B14). This is reasonable and even expected  given that those ``canonical" parameters were obtained within the standard disk-like model whose propagation characteristics are quite different from those of the spiral arm model that we consider here.

An interesting related phenomenon is the unexpected energy dependence of the positron fraction, $e^+/(e^++e^-)$, of CRs. Since positrons in standard scenarios are secondary particles, the ratio is expected to decrease with energy \citep{GALPROPelectron}. However, the {\sc pamela} satellite measurements revealed that the ratio increases with the energy above 10~GeV up to at least 100~GeV (\citealt{Adriani}). More recent measurements by AMS-02 show leveling at 300~GeV \citep{AMSpositrons}. 

Given that standard models could not explain this behavior, several solutions were suggested. The increased positron fraction at higher energies can be explained in any scenario having an additional primary population of pairs with a hard spectrum, such that it dominates the population of secondary positrons above 10~GeV. Pairs can naturally be created by dark matter annihilation \citep{DarkMetter1,DarkMetter2}, by pulsars (\citealt{Pulsar1}, \citealt{Pulsar2}, \citealt{Pulsar3}, \citealt{Pulsar4}, and \citealt{Pulsar5}) or in aged SNRs (\citealt{Positrons-secondaries1}, \citealt{Positrons-secondaries2}, \citealt{Positrons-secondaries3}, and \citealt{Positrons-secondaries4}).

Another type of explanation directly related to the PLD was proposed by \cite{Pamela}. If the CR sources are concentrated at a finite distance from the solar system, as is expected from the spiral arm structure, the paucity of short path lengths implies that primary electrons with a high enough energy will cool before reaching the solar system. In contrast, the secondary positrons are formed by protons that effectively do not cool, such that they can be formed in the solar system's vicinity. For the local electron cooling rate determined by Synchrotron and inverse-Compton scattering, and the typical CR age determined by Beryllium isotope ratios, this behavior predicts that the Positron fraction should start increasing above 10~GeV, thus explaining the ``Pamela Anomaly", effectively without any free parameters.

In the present work we study the B/C and sub-Fe/Fe ratios using an extended version of the B14 model, as is described in \S\ref{sec:upgrade}.
One of the major uncertainties arising when studying the sub-Iron to Iron ratio and its implications to the PLD, is the energy dependent spallation cross-sections. Since many of the partial cross-sections are poorly measured, the values used are often the results of fits and extrapolations, giving rise to large uncertainties \citep{Moskalenkocross,WebberSoutoulcross}. The problem is aggravated below a few GeV/nuc., where the cross-sections have larger energy dependences \citep{Moskalenkolightcross,Schwallercross}, and for heavier elements (e.g., see appendix II in \citealt{GarciaMunoz,Titarenkocross,Sistersoncross}). 
We limit the analysis to energies between 10~GeV/nuc. to 1000~GeV/nuc. Although it is not clear by how much it decreases the uncertainties in the cross-sections, the weaker energy dependence will translate into a smaller uncertainty in the inferred spectral slope of the diffusion coefficient. We use the nominal model parameters of B14, as summarized in \S\ref{sec:parameters}, and search the optimal diffusion coefficient spectral index and normalization, namely, $\delta$, and $D_\mathrm{0}$ in $D = D_0 (E/E_0)^\delta$ \footnote{Throughout the paper, the term ``diffusion coefficient" actually refers to the normalization $D_\mathrm{0}$ and not $D(E)$.}, with $E_0 = 3$ GeV/nuc., to fit both the B/C from AMS-02 \citep{AMS-B/C} and the sub-Fe/Fe data from HEAO-3 \citep{BinnsHEAO} and SANRIKU \citep{SANRIKU}. We carry out the analysis twice, once for a disk-like model having no spiral arms, and once for the spiral arm model of B14.

We begin in \S\ref{sec:model} by briefly describing the model we developed in B14 and the parameters we use. In \S\ref{sec:upgrade} we detail the improvements carried out for the present work. The results are then described \S\ref{sec:results}, the main part of which includes  separate fits of $\delta$, and  $D_{0}$ to the observed B/C ratio and to the sub-Fe/Fe ratio. The implications of these results are then discussed in \S\ref{sec:discussion}.

\section{The numerical model}
\label{sec:model}

In B14 we developed a CR advection-diffusion model with which we recovered the B/C ratio. The nuclear network in the model included all stable and long lived radioactive isotopes between Beryllium and Oxygen. The model followed the propagation of primary and secondary CRs from their sources, primarily located on spiral arms.  
Unlike present day state of the art simulations (such as {\sc galprop}, \citealt{StrongReview} and {\sc dragon}, \citealt{Dragon}) that solve the partial differential equations of the describing advection and diffusion, our model uses a Monte Carlo (MC) algorithm.

Instead of simulating the small steps corresponding to the physical mean free path (m.f.p.) of the CRs, around 1~pc, our MC simulation uses a larger {\em effective} m.f.p., which lumps together many small effective physical steps. This corresponds to a larger effective time step chosen to be $\tau_\mathrm{esc}/100$, where $\tau_{esc}\approx {Z_\mathrm{h}^2 / 2 D}$ is the typical escape time of a CR from the galaxy\footnote{Although the time step $\delta t(E)$ is energy dependent, the effective m.f.p. is not, as it depends on the combination $\delta t(E) \times D(E)$.}. In the present simulations, we fix $Z_\mathrm{h}$ to be $250$~pc, like our nominal model in B14. This corresponds to an effective m.f.p.\ of 25~pc, Which is smaller than the typical length scale over which the gas density varies.

Each time step, we check whether the CR bundle had left the galaxy. If it did not, we calculate the grammage that the bundle traversed in this time step. We then calculate the probability that the particles had undergone a nuclear reaction with ISM. If the particles did undergo spallation into secondary particles, we follow the latter with the same methodology. We nominally use the {\sc yieldx} subroutine of \cite{XSection} for the nuclear cross-secitons. However, we repeat the analysis with \cite{WebberSoutoulcross} to obtain a handle on the uncertainty introduced by the nuclear cross-sections, as described below. For long-lived radioactive isotopes, we also check each time step for radioactive decay. The code also considers Coulomb and ionization losses as elaborated in \cite{Mannheim1994} and applied in {\sc galprop} \citep{StrongNucleons}.

For the present simulations, each run uses $10^{9}$ CR bundles in the spiral arms simulations and $10^{8}$ in the disk-like simulations. For each bundle, we randomly choose its initial isotopic identity and initial location in the galaxy, 
 and follow it until it escapes the galaxy, cool below the $10$~GeV/nuc., break into isotopes lighter than the lightest element simulated or reach the solar vicinity. In the latter case 
we record all simulated isotopes (Beryllium to Silicon in the light elements simulations and Scandium to Nickel in the heavy elements simulations), in a 3D array (Z, A, energy). The energies are recorded in 10 bins per decade, covering the range of 10~GeV/nuc. to 1000~GeV/nuc..

\subsection{The extended code}
\label{sec:upgrade}

Below we detail specific modifications and improvements of the code used in this work. 

\subsubsection{Nuclear network}
\label{sec:nuclear}

As we are interested in second order corrections to the B/C ratio, we extended the code to describe the spallation network all the way up to Silicon.

In addition to the range of ``light" elements necessary for the aforementioned calculation, we are also interested in calculating the sub-Fe/Fe ratio. We therefore added the nuclear network of heavier elements, from Scandium to Nickel as well, using the {\sc yieldx} \citep{XSection}, or alternatively  \cite{WebberSoutoulcross} for the partial spallation cross-sections describing the interaction with Hydrogen. We use the approximate formulae of \cite{Karol1975} for the total inelastic cross-sections of interactions with Helium.

\subsubsection{Faster code}
\label{sec:fast}

The code was accelerated by limiting the CR diffusion in the spiral-arms models to take place only within of a radius 5 times the halo size around  the solar system. This captures 97\% of the total CRs reaching the solar system, when comparing to a full calculation without the distance cutoff, but increases the speed by about an order of magnitude to reach similar statistics.
For the disk-like model simulations, this is unnecessary. The azimuthal symmetry allows for much better statistics than the spiral-arms simulation with much fewer CR bundles.

In order to increase the accuracy, we decrease the time step 10-fold when the CR bundle is within a distance of  150~pc from the solar system, an additional 10-fold when the bundle is within 120~pc, and we take a time step for which the effective mean free path (m.f.p.) is of order the actual physical m.f.p. (of order 1~pc), when the CR bundle is within 100~pc from the solar system. 

\subsection{Initial composition}
\label{sec:initial}

The initial composition for which we obtain the optimal fit to the measured secondary/primary isotope ratios in the light elements simulations is 37\% Carbon (by number), 3\% Nitrogen, 52\% Oxygen, 4\% Neon, 2\% Magnesium and 2\% Silicon. Note that although the method for finding the initial composition is the same as in B14, the composition is somewhat different because the code now follows more isotopes, as described in \S\ref{sec:nuclear}.
For the heavy elements simulations, the best fit to the observed Ni/Fe ratio, is obtained for an initial composition consisting of 95\% Iron and 5\% Nickel.



\subsection{The nominal model parameters}
\label{sec:parameters}

Table.\ \ref{table:parameters}  summarizes the nominal model parameters obtained by or referenced within B14. These parameters are separately fixed for either the spiral-arms simulations or the disk-like simulations.

\begin{table}[h]
\centering
\caption{Nominal Model Parameters}
\begin{tabular}{ c c c c c c c c}
\hline
 \multirow{3}{*}{parameter} & \multirow{3}{*}{definition} & value for & value for\\
   & & spiral arm & disk-like\\
   & & model & model\\
\hline
 $Z_\mathrm{h}$ & Half halo height & 250 ~pc & 1kpc\\
$\tau_\mathrm{arm}$ & Last spiral arm passage & 5 ~Myr & \\ 
$i_4$ & 4-arms set's pitch angle & 28$^\circ$ & \\
$i_2$ & 2-arms set's pitch angle & $11^{\circ}$ & \\
\multirow{2}{*}{$\Omega_4$} & Angular velocity of & \multirow{2}{*}{15 (km/s) kpc$^{-1}$} & \multirow{2}{*}{} \\
 & the 4-arms set & & \\
\multirow{2}{*}{$\Omega_2$} & Angular velocity of & \multirow{2}{*}{25 (km/s) kpc$^{-1}$} & \multirow{2}{*}{} \\
 & the 2-arms set & & \\
\multirow{2}{*}{$f_\mathrm{SN,4}$} & Percentage of SN in & \multirow{2}{*}{48.4\%} & \multirow{2}{*}{} \\
 & the 4-arms set & & \\
\multirow{2}{*}{$f_\mathrm{SN,2}$} & Percentage of SN in & \multirow{2}{*}{24.2\%} & \multirow{2}{*}{} \\
 & the 2-arms set & & \\
\multirow{2}{*}{$f_\mathrm{SN,CC}$} & Percentage of core collapse & \multirow{2}{*}{8.1\%} & \multirow{2}{*}{80.7\%} \\
 & SNe in the disk & & \\
\multirow{2}{*}{$f_\mathrm{SN,Ia}$} & Percentage of  & \multirow{2}{*}{19.3\%} & \multirow{2}{*}{19.3\%} \\
 & SN Type Ia & & \\
\hline
\end{tabular}
\label{table:parameters}
\end{table}

\section{Observational Datasets}

We compare the model predictions for the B/C ratio with the AMS-02 results, whose  advantage over previous datasets is the notably better statistics.  In total, there are 10 data points between 10 GeV/nuc. and 1000 GeV/nuc.  \citep{AMS-B/C}.

The sub-Fe/Fe datasets used both the HEAO-3  \citep{BinnsHEAO} and the SANRIKU measurements \citep{SANRIKU}. In total, there are 21 data points between 10 GeV/nuc.\ and 1000 GeV/nucleon.

\section{Results}
\label{sec:results}

We have separately fitted the B/C data and the sub-Fe/Fe data to 
the two CR source models: the  spiral {arms} sources, and a disk-like, azimuthally symmetric, one. For each model we then checked its consistency, as the different datasets (B/C and sub-Fe/Fe) should be fitted with the same models parameters. If we find consistency the optimal parameters then teach us about CR diffusion in the interstellar medium.

To obtain the best fit values, we minimize the $\chi^2$ for each simulation. The uncertainty used in estimating the $\chi^2$ values include both those arising from the (Monte Carlo) simulation and the quoted observational errors which are typically an order of magnitude larger.

%
%
%
%
%
%

\subsection{B/C}
\label{sec:BC}

The fit to the optimal spectral slope $\delta$ and diffusion coefficient $D_0$ is carried by calculating the $\chi^2$ over a two dimensional array of values. Then, the set of $\chi^2$ for each value of $\delta$ in the range of 0.3 to 0.7 is fitted with a polynomial, and the minimum found. 

The optimal diffusion coefficient as a function of $\delta$ is given in table \ref{table:combinations}. By comparing the disk-like models to the spiral arm models, it is evident that the typical $D_0$ required to fit the observations in ``standard" disk-like models is much larger than the typical $D_0$ required to fit the same data in the spiral arm model. This recaptures the result previously described in B14, and arises from the smaller halo in the spiral arm model. 

\begin{figure}[h]
\centerline{\includegraphics[width=\figwidth]{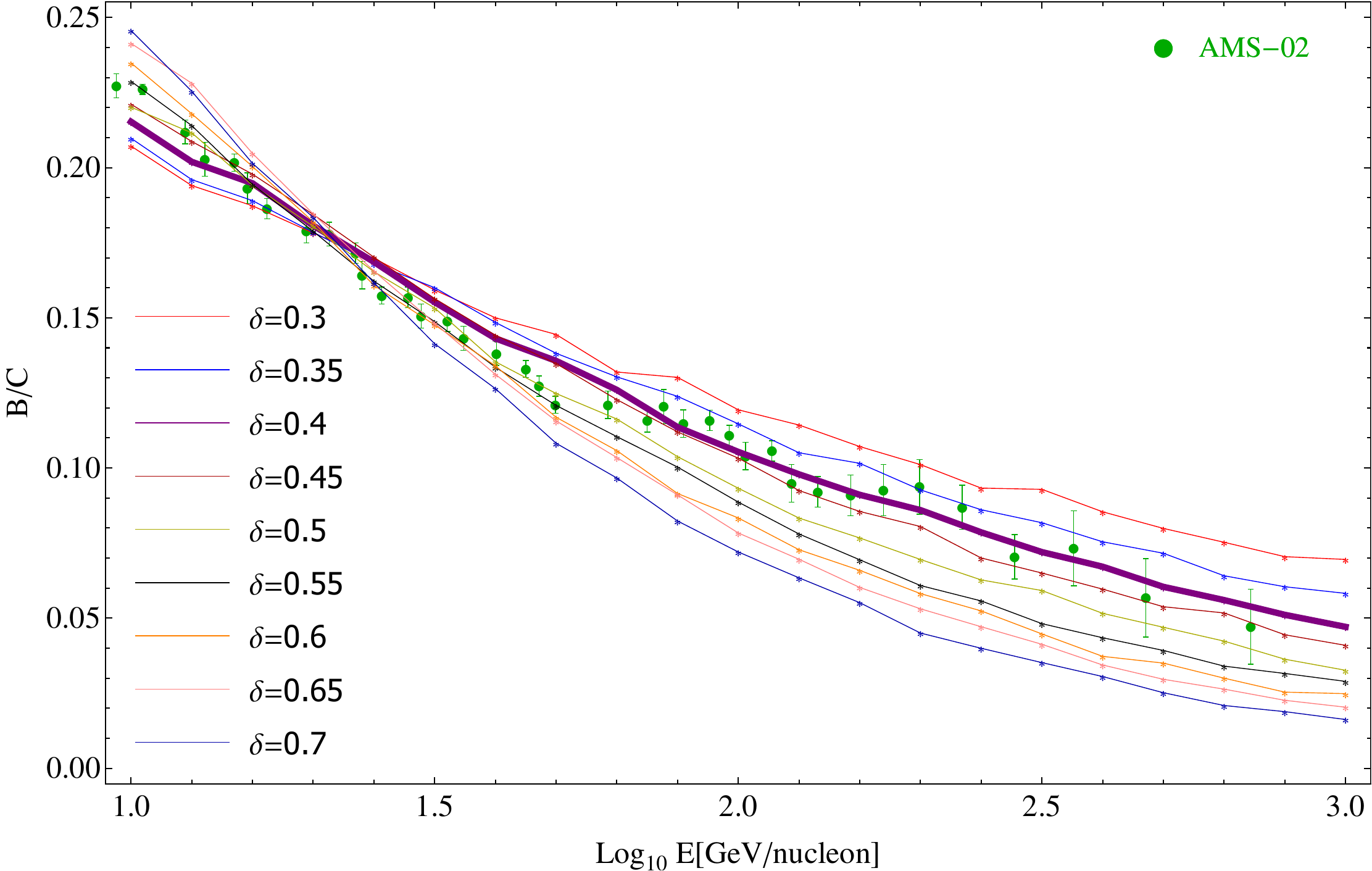} }
\caption{The optimal $\chi^2$ fit of the B/C data for several $\delta$ values in the range  0.3 to 0.7, for disk-like simulations. Data is  taken from \cite{AMS-B/C}. The thicker purple line corresponds to the overall best fit, with $\delta=0.45$.}
\label{fig:BC_slopes_disk}
\end{figure}

\begin{figure}[h]
\centerline{\includegraphics[width=\figwidth]{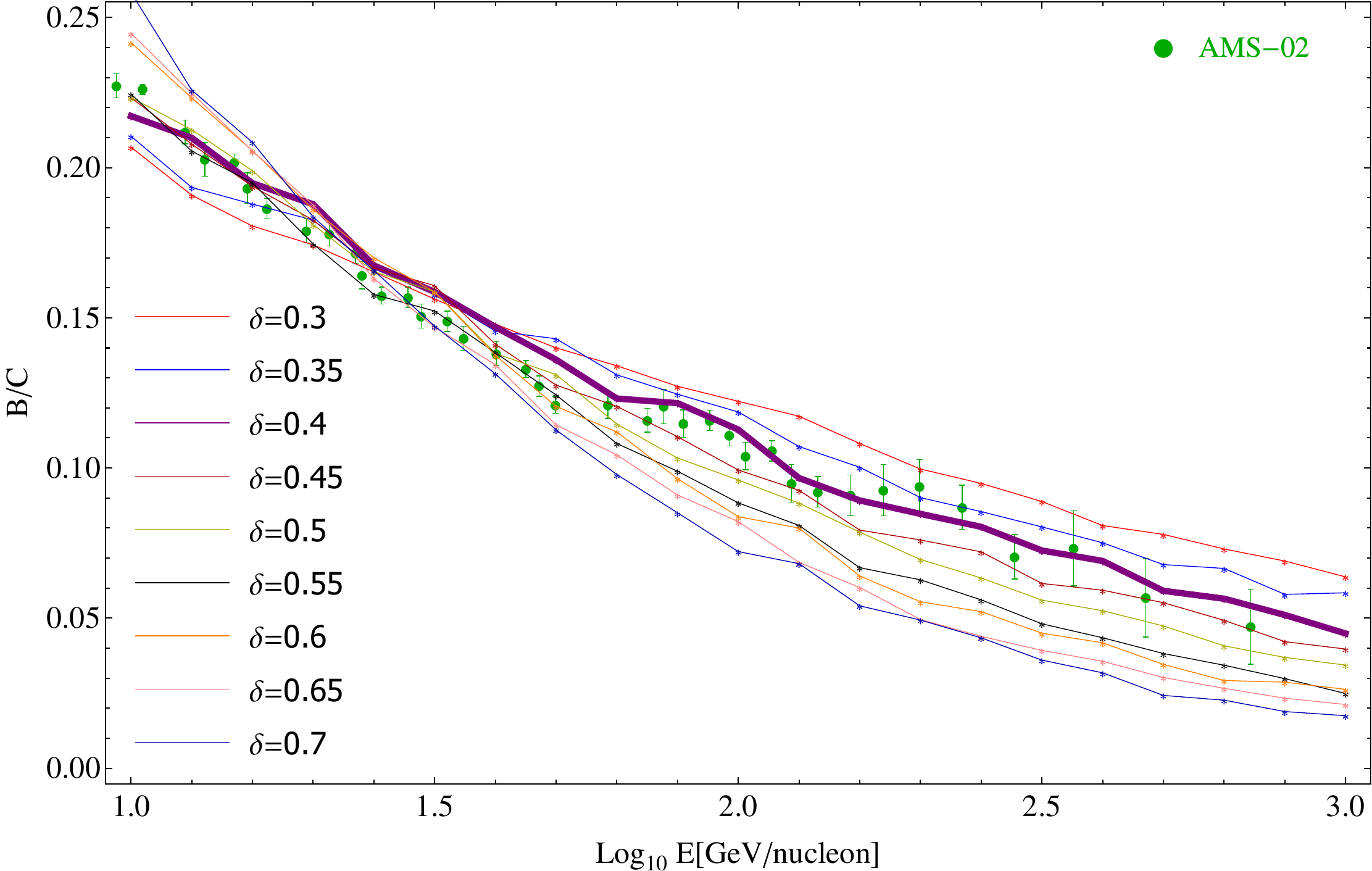} }
\caption{Similar to fig.\ \ref{fig:BC_slopes_disk} but for the spiral arm model. The overall optimal fit here is also $\delta=0.45$.}
\label{fig:BC_slopes_spiral}
\end{figure}

The model fits to the data, as a function of $\delta$, are depicted in fig.\ \ref{fig:BC_slopes_disk} for the disk-like model and fig.\ \ref{fig:BC_slopes_spiral} for the spiral arm model. One sees from these fits that one requires roughly the same $\delta\sim 0.4$ (thicker purple line) to fit both the disk-like and spiral arm models.

Figs.\ \ref{fig:contour_disk} \& \ref{fig:contour_spiral} depict the $\chi^2$ as a function of both $\delta$ and $D_0$. 

The latter plots also demonstrate that for a given value of $\delta$ there is a narrow range of $D_0$ for which there is a reasonable fit. However, without prior knowledge of $\delta$, there is a much wider allowed range for optimal $D_0$.    

\subsection{Iron}
\label{sec:Iron}

\begin{figure}[h]
\centerline{\includegraphics[width=\figwidth]{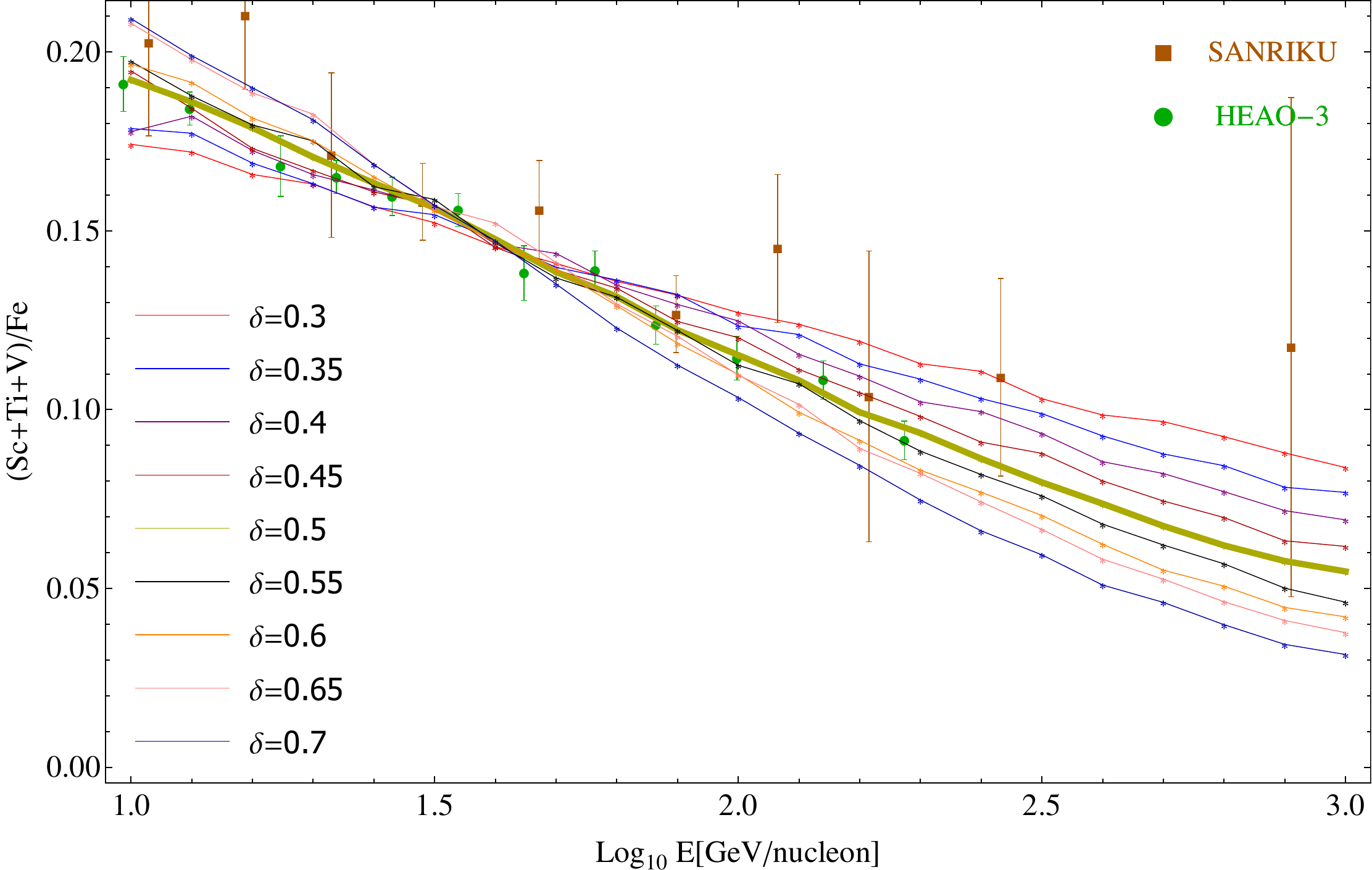} }
\caption{The optimal $\chi^2$ fit of the sub-Fe/Fe data for a set of $\delta$'s between 0.3 to 0.7, in disk-like simulations. The thicker mustard colored line corresponds to the overall best fit, with $\delta=0.5$. Data taken from HEAO-3 \citep{BinnsHEAO} and SANRIKU \citep{SANRIKU}.}
\label{fig:Iron_slopes_disk}
\end{figure}

\begin{figure}[h]
\centerline{\includegraphics[width=\figwidth]{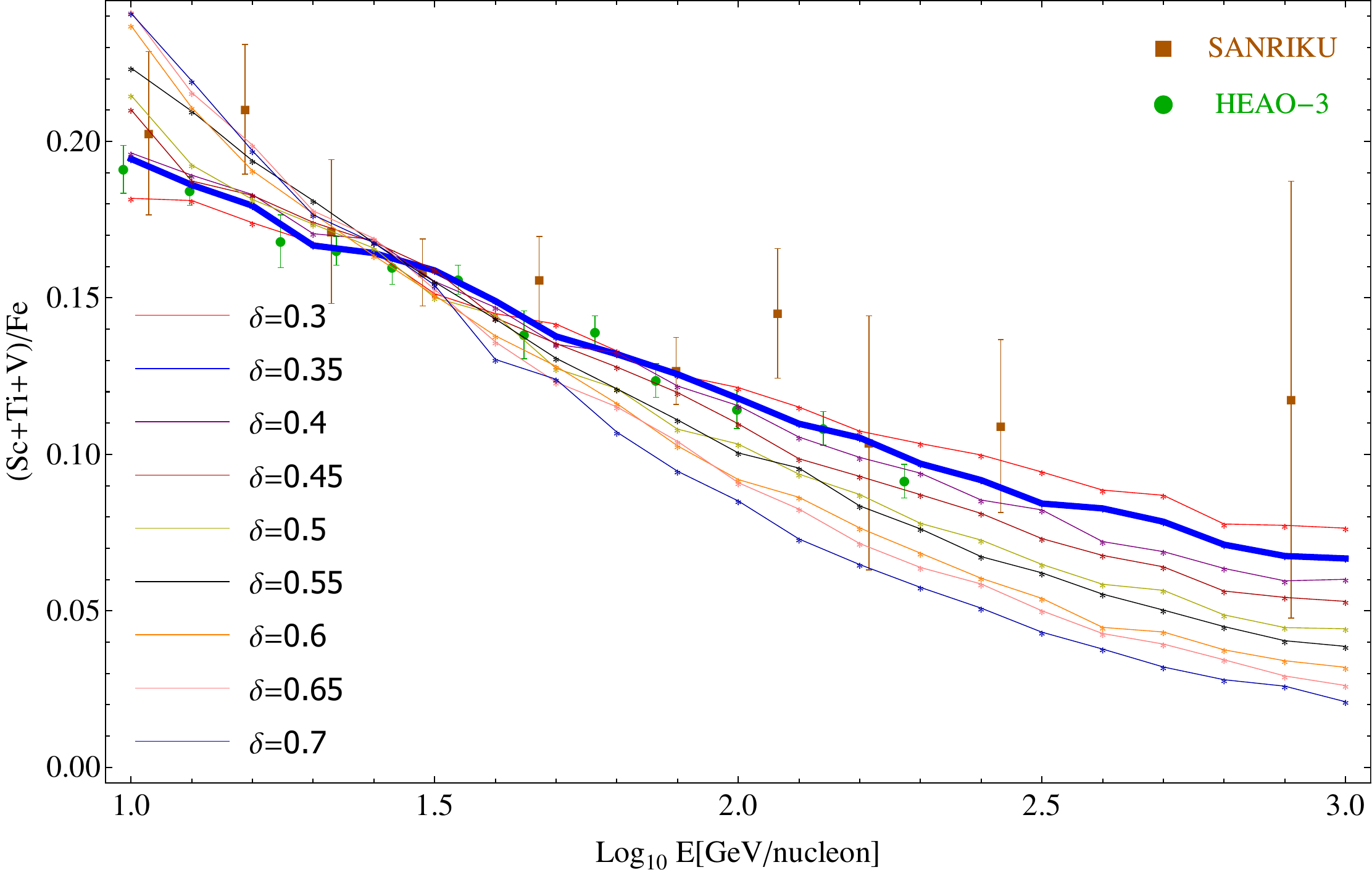} }
\caption{Similar to fig.\ \ref{fig:Iron_slopes_disk} but for the spiral arm model. Unlike the disk-like model, the overall optimal fit here is $\delta=0.35$ depicted a thick blue line.}
\label{fig:Iron_slopes_spiral}
\end{figure}

Independently of fitting the B/C data, we fit the sub-Iron to Iron data in a similar way. The model fits to the Iron data as a function of $\delta$, are depicted in fig.\ \ref{fig:Iron_slopes_disk} for the disk-like model and fig.\ \ref{fig:Iron_slopes_spiral} for the spiral arm model. Unlike the B/C fits, here the optimal fits require a different $\delta$ values for the two models. For the disk-like model, the optimal is $\delta\sim 0.5$ while it is $\delta\sim 0.35$ for the spiral arm model. The source of the difference points to an interesting physical effect associated with the different path length distribution. Because the cross-section of the heavier elements is larger, their mean free paths are much smaller relative to the distance traversed,
thus giving rise to saturation (especially at lower energies), whereby an increase in the grammage does not increase as much the ratio between secondary and primary isotopes. This saturation causes a decreases in the secondary/primary slope. However, because of the different path length distribution, the effect is notably larger for the disk like model with an exponential PLD. Thus, in order to compensate for this effect, one requires a steeper diffusion power law index to explain the small observed slope. For the spiral arm model, the saturation is less important such that the diffusion power law index should be closer to the observed sub-Fe/Fe slope.   

Figs.\ \ref{fig:contour_disk} \& \ref{fig:contour_spiral} show the $\chi^2$ as a function of both $\delta$ and $D_0$.

\subsection{B/C vs. sub-Fe/Fe comparison}
\label{sec:Contour}

\begin{figure}[h]
\centerline{\includegraphics[width=\figwidth]{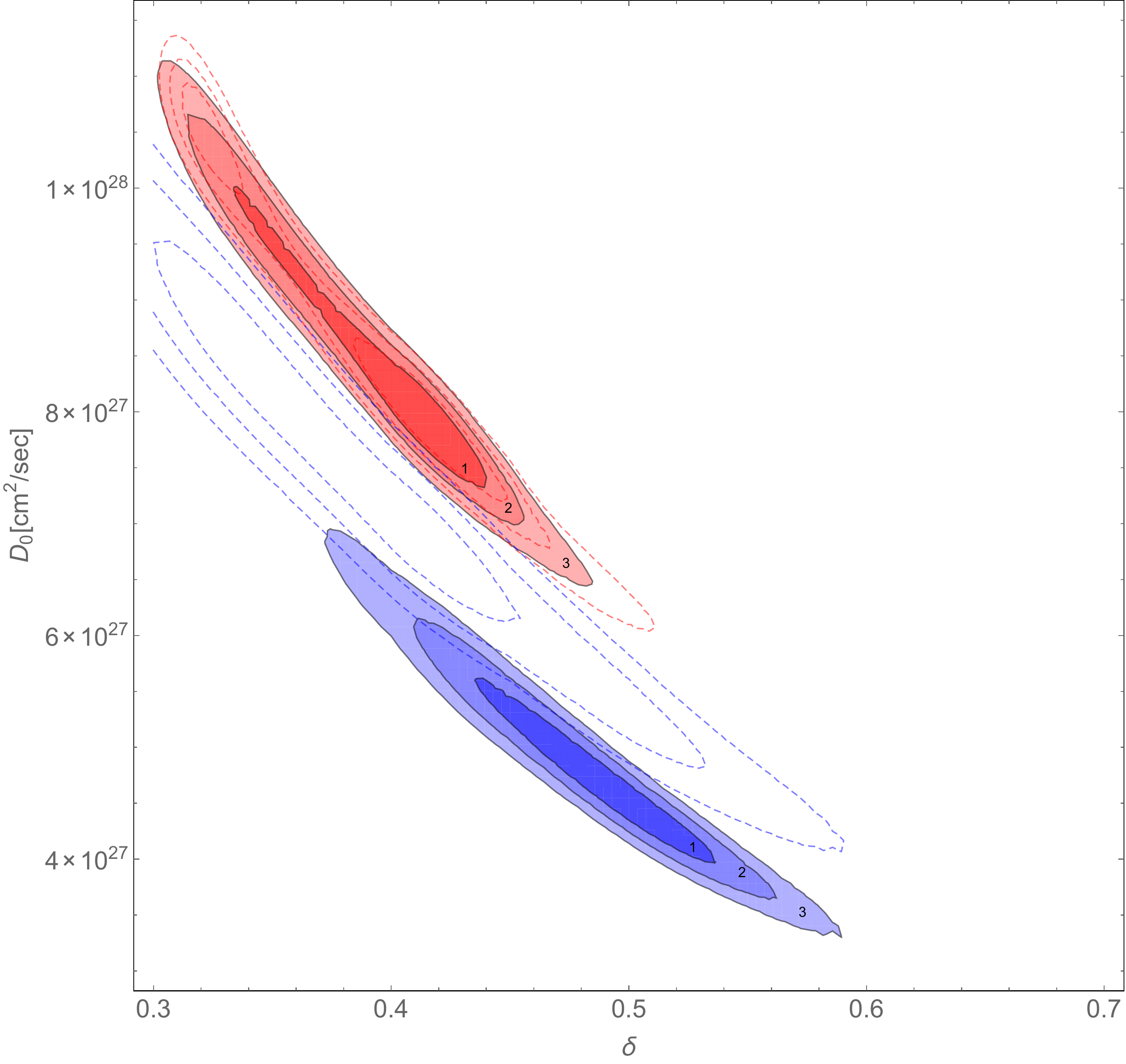} }
\caption{A contour plot of $\chi^2$ for the disk-like model. The red contours correspond to the B/C fit. The blue contours correspond to the sub-Fe/Fe fit. The dashed lines correspond to the same contours as obtained when replacing the cross-sections of \cite{XSection} with those compiled by \cite{WebberSoutoulcross}. Note that the discrepancy between the B/C and the sub-Fe/Fe derived model parameters decreases, but still remains.   
}
\label{fig:contour_disk}
\end{figure}

\begin{figure}[h]
\centerline{\includegraphics[width=\figwidth]{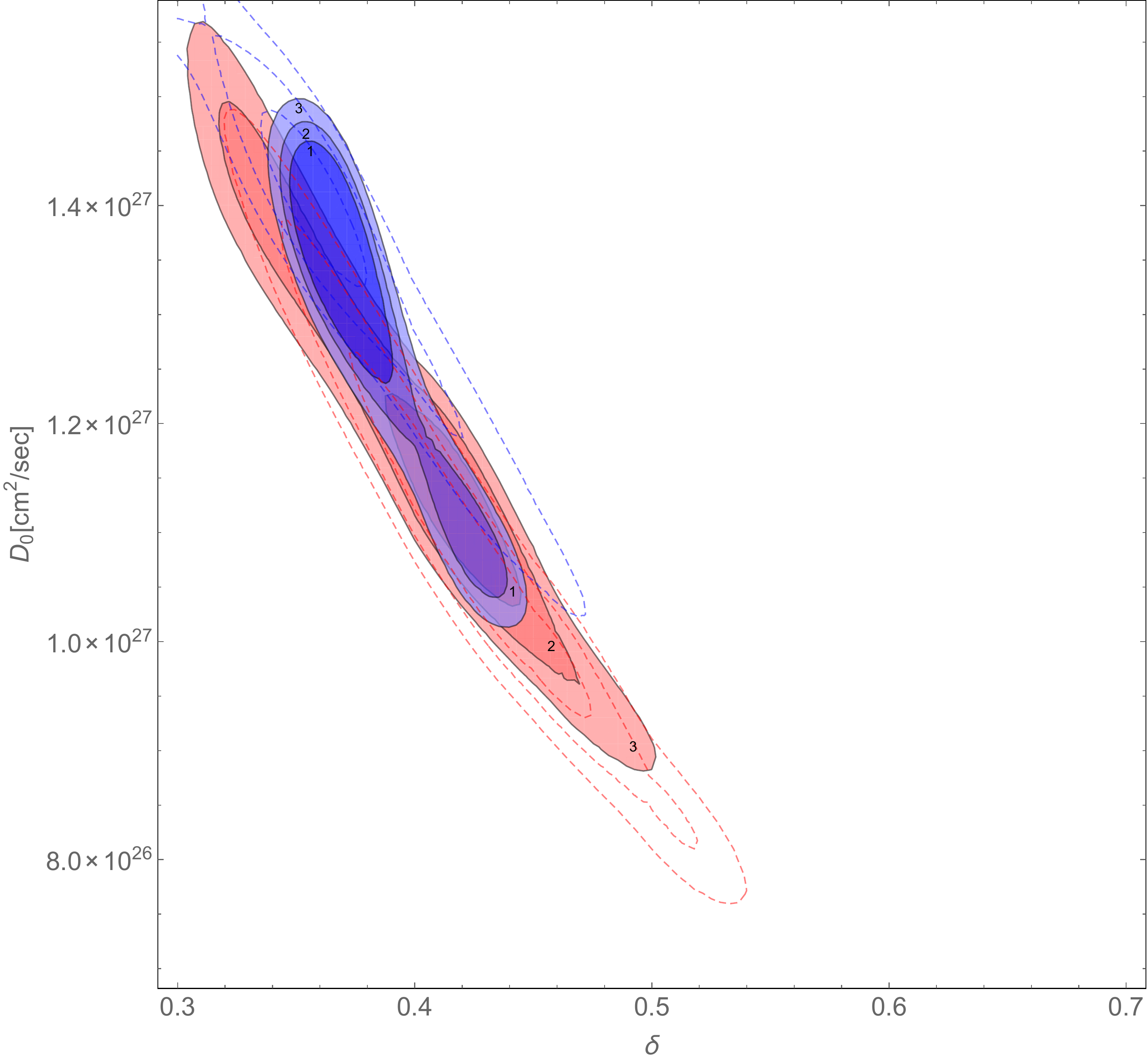} }
\caption{A contour plot of  $\chi^2$ for the spiral arm model. The red contours correspond to the B/C fit. The blue contours correspond to the sub-Fe/Fe fit . The dashed contours are similar to those in fig.\ \ref{fig:contour_disk}.
}
\label{fig:contour_spiral}
\end{figure}

The above analysis was carried out separately with B/C and sub-Fe/Fe data sets. However, any consistent model should be able to fit both datasets with the same model parameters. This implies that the $\chi^2$ contours depicted in figs.\ \ref{fig:contour_disk} \& \ref{fig:contour_spiral} should overlay each other. 
A quick inspection reveals that only the spiral arm model gives consistent parameters, with $D_0 \sim 10^{27}$cm$^2$s$^{-1}$ to $1.2\times 10^{27}$cm$^2$s$^{-1}$ and $\delta \sim 0.35-0.45$. This can also be seen in table \ref{table:combinations}. While the disk-like model requires a similar $\delta$ to fit both datasets, the iron data requires a diffusion coefficient that is of order 35\% smaller from the diffusion coefficient required to recover the B/C data, for any given $\delta$. The smaller diffusion coefficient corresponds to a larger grammage. This recaptures the claims raised by \cite{GarciaMunoz}, that the sub-Fe/Fe data requires a larger grammage than the B/C data requires. Here we limit ourselves to high energies ($>$10 GeV/nuc.) where we expect smaller uncertainties in the energy dependence of the cross-sections, and reach the same conclusion. 

This inconsistency is manifested in the minimum $\chi^2/\mathrm{d.o.f.}$ (per degree of freedom) obtained in the combined analysis. For the disk-like model, we find  $\chi^2/\mathrm{d.o.f.}= 3.66$ with $\mathrm{d.o.f.} = 31-2=29$. However, for the spiral arm model it is $\chi^2/\mathrm{d.o.f.}= 0.95$, for the same number of degrees of freedom. Formally, this implies that the disk like model can be ruled out at the $10^{-10}$ level. 

To see how sensitive this conclusion is to the nuclear cross-sections employed, we repeated the analysis with the cross-sections of \cite{XSection} replaced with those of \cite{WebberSoutoulcross}. The results are depicted as the dashed contours in figs. \ref{fig:contour_disk} and \ref{fig:contour_spiral}. Evidently, the discrepancy remains though it is more than twice smaller. In other words, the uncertainties in the cross-sections are sufficiently large that one cannot unequivocally use the B/C and sub-Fe/Fe data to rule out a disk-like model, though it is inconsistent with the present spallation cross-sections.

One should emphasize that the actual values of the diffusion coefficients are less certain than from these fits. This is because we set the nominal halo sizes for the two models. Changing the halo sizes would also scale the values of the optimal $D_0$ obtained. While this introduces an uncertainty in the actual values of $D_0$ this will not resolve the inconsistency between the two data sets in the disk-like models and will not introduce any inconsistency for the spiral arm model. That is, the scaling would be the same for both the B/C and sub-Fe/Fe such that the consistency (in the spiral arm model) or inconsistency (in the disk-like model) between the dataset fits would remain. 

\begin{table}[h]
\centering
\caption{The optimal $D_\mathrm{0}$ in units of $10^{27}$cm$^2$s$^{-1}$ as obtained for the two models, when separately fitting the B/C and the sub-Fe/Fe data.}
\begin{tabular}{c c c c c c c c c}
\hline
& \multicolumn{2}{c}{$D_\mathrm{0}$ from spiral arm model} &   \multicolumn{2}{c}{$D_\mathrm{0}$ from disk-like model} & \\
\hline 
$\delta$ &  B/C &  sub-Fe/Fe &  B/C &  sub-Fe/Fe & \\
\hline 0.3   &   1.6   &   1.65     &   10.9      &   8.8   \\
  0.35   &   1.37      &   1.4      &   9.5      &   7.4   \\ 
  0.4   &   1.17      &   1.2      &   8.2      &   6.2   \\
  0.45   &   1.03      &   1.01      &   7      &   5.3   \\
  0.5   &   0.89      &   0.9      &   6.2      &   4.5   \\
  0.55   &   0.8      &   0.74      &   5.5      &   3.8   \\
  0.6   &   0.65      &   0.66      &   4.7      &   3.2   \\
  0.65   &   0.58      &   0.56      &   4      &   2.6   \\
  0.7   &   0.5      &   0.5      &   3.6      &   2.3   \\
  \hline
\end{tabular}
\label{table:combinations}
\end{table}
 
  \eject
 
\section{discussion \& Summary}
\label{sec:discussion}

In previous work, we developed a fully 3D CR diffusion model which not only considers that most CR acceleration takes place in the vicinity of spiral arms, but also that these arms are dynamic \citep{BoverC}. It was shown that by adding these observationally based components to the model,  one recovers  observed secondary to primary ratios, such as of Boron to Carbon. In particular, it was found that the ratio increases with energy below 1~GeV/nuc. and decreases above this energy, instead of the monotonic decrease with energy expected in the simplest galactic diffusion model (e.g., \citealt{CRreview}, without having to assume additional assumptions, such as reacceleration or a galactic wind). This is because the age of the CRs at low energies is not determined by the diffusion time from the spiral arms, but instead governed by the (energy independent) time since the last spiral arm passage. Since below 1~GeV/nuc., the particles are non-relativistic, a fixed age translates into a grammage which is increasing with energy, and correspondingly an increasing secondary to primary ratio at low energies. 

One very important aspect of this model is that the path length distribution is different from the one found in standard diffusion models. In the latter, the PLD is typically close to being exponential. However, if most CRs arrive from a distance, then the PLD will show a paucity of small path lengths (compare fig. 4 to fig.6 in B14).

The different PLD has an interesting ramification. CRs arriving after having passed a short path length necessarily remain close to the galactic plane. In contrast, CRs having a long path length could stray further from the plane before returning back. Since the ISM density falls with the distance from the plane, CRs with short paths therefore experience a higher average ISM density than CRs with long paths. Thus, a distribution which is missing the short path lengths, as is the case in the spiral arm model, will have a lower average interaction with the ISM for a given average path length. In other words, the average grammage will be lower for the same average physical length. This result implies that if the spiral arm model is to recover the observed secondary to primary ratio, the model has to keep the CRs closer to the galactic plane where the density is higher. For this reason, the typical halo size required to fit the same secondary to primary ratio data is lower (typically a few hundred pc, compared with the 1 to 4 kpc in more standard diffusion models). However, because the typical age is closely related to the ratio between the radioactive and stable Beryllium isotopes, which is a model independent observation, the smaller halo requires a smaller diffusion coefficient. At 1~GeV/nuc. it should be of order $10^{27}$ and not $10^{28}$cm$^2$s$^{-1}$.

Here we extended the model developed in B14, as detailed in \S\ref{sec:upgrade}. In particular, we enlarged the spallation network to describe the spallation of Iron and Nickel as well as their spallation products. Thus, we could predict the sub-Fe/Fe ratio as a function of model parameters, in addition to the B/C ratio. We carried a parameter study whereby we fit both the spectral slope $\delta$ of the diffusion coefficient and its normalization $D_0$ (while keeping the other model parameters fixed at their nominal values). This was carried out twice, for a (``standard") disk-like model and for a spiral arm model, in the energy range of 10~GeV/nuc. - 1000~GeV/nuc.. The reason for focusing on high energies is to avoid the notorious uncertainties associated with the spallation cross-sections at lower energies \citep[e.g.,][]{Moskalenkocross,WebberSoutoulcross}. These uncertainties become more acute as $Z$ increases  (see \citealt{Titarenkocross,Sistersoncross} and also Appendix II in \citealt{GarciaMunoz}).

In a standard disk-like model, we find an inconsistency between the diffusion coefficient required to fit the B/C and the value required to fit the sub-Fe/Fe ratio.  Phrased differently, it implies that the sub-Fe/Fe ratio requires more grammage than the B/C ratio. One way to resolve the inconsistency present in the standard disk-like model would be to modify the cross-sections at the higher energies over which the fit to observational data was carried out. This is not inconceivable given that by replacing the cross-sections of \cite{XSection} with those of \cite{WebberSoutoulcross}, we reduce the discrepancy by more than a factor of 2.

Another way to resolve the inconsistency within the standard disk like model would be to truncate the short paths lengths by letting the CRs interact with material at the SNR. This explanation which was originally proposed to resolve the PAMELA anomaly \citep{Blasi2009}, would requires however additional free parameters. A priori it is not clear that the parameters that resolve the PAMELA anomaly would also resolve this inconsistency.  

A spiral arm source distribution consistently requires the same diffusion coefficient (or grammage) when fitting the sub-Fe/Fe and B/C datasets. This arises because the average mean free path  for spallation of Iron is notably smaller than that of Carbon and Oxygen, such that the sub-Fe/Fe and the B/C are not sensitive to the same path lengths. As a consequence, modifying the PLD in the spiral arm model doesn't change the average grammage needed to explain the sub-Fe/Fe and the B/C by the same factor.  

  The best fit values are a diffusion spectral slope of $\delta=0.35$ to $\delta=0.43$, and a diffusion coefficient normalization between $D_\mathrm{0}=1\times10^{27}$cm$^{2}$s$^{-1}$ and $D_\mathrm{0}=1.5\times10^{27}$cm$^{2}$s$^{-1}$ (though the diffusion coefficient range would scale with the halo size, which was chosen here to be 250 pc). To summarize, the consistency found here is one of several successful predictions borne from the spiral arm model: 
\begin{enumerate}

\item The increase in the positron fraction above $\sim 10$~GeV is a necessary outcome given that the primary electrons which arrive from a finite distance cool through inverse-Compton and synchrotron radiation while secondary positrons can form locally \citep{Pamela,Gaggero2013}. 
\item The increase in the B/C ratio for $E/\mathrm{nucleon} \lesssim 1$~GeV/nuc. is recovered since the CR age saturates over low energies at the time since the last spiral arm passage (B14).
\item The concentration of CR sources around the galactic spiral arms and lower diffusion coefficient give rise to a temporal variation in the CR flux, seen as 145 Myr cycles in the CR exposure ages behavior of iron meteorites \citep{ShavivNewAstronomy}. 
\item The smaller halo of the spiral arm model explains the 32 Myr oscillation seen in the 550 Myr long paleoclimatic record \citep{Shaviv2014}.
\item The lower diffusion coefficient implies that the expected anisotropy in the arrival direction of cosmic rays should be smaller by an order of magnitude from predictions of disk-like models with a larger diffusion coefficient, and consistent with the measurements between 1 to 100 TeV \citep{ShavivAnisotropyInPreparationOrSubmitted}. 
\item The arm dynamics are responsible for a CR ``wake" behind the arms with softer CRs, which give rise to a softer $\pi^0$ produced spectrum. Together with the smaller halo, the predicted spectral slope map in $\gamma$-rays at around 1-10 GeV is consistent with the FERMI observations \citep{NavaGammaRaysInPreparationOrSubmitted}. 
\item Last, as we have shown here, the inconsistency between the required diffusion coefficient to fit the B/C and the sub-Fe/Fe ratios is resolved given the different Path Length Distribution. 
\end{enumerate}

Thus, the present results introduces further circumstantial evidence to a list that strongly implies that CRs source inhomogeneity plays an important role in CR physics.

\section*{Acknowledgements}

This work is has been supported by ISF grant no.\ 1423/15 (DB \& NJS), as well as by the I-CORE Program of the Planning and Budgeting Committee and The Israel Science Foundation (1829/12). NJS also gratefully acknowledges the support of the IBM Einstein Fellowship at the Institute for Advanced Study.


\def\jcap{J.\ Cos.\ Astropart.\ Phys.}
\def\na{nature}

\bibliography{Iron.bib}

\end{document}